\documentclass{PoS}
\usepackage{subfig,amsmath}

\title{Mass Anomalous Dimension at Large N}

\ShortTitle{Mass Anomalous Dimension at Large N}

\author{\speaker{Liam Keegan}\\
       Instituto de F\'{i}sica Te\'{o}rica UAM/CSIC, Universidad Aut\'{o}noma de Madrid, Cantoblanco, 28049 Madrid, Spain.\\
       E-mail: \email{liam.keegan@uam.es}}

\abstract{
In this work we attempt to determine the mass anomalous dimension of the SU(N) gauge theory with two Dirac fermions in the adjoint representation, in the limit of large N. The method uses the mode number of the Dirac operator, as done in Ref.~\cite{Patella:2012da} for the SU(2) theory in a large volume. Taking advantage of large--N volume reduction we do this on a $2^4$ lattice, but we should still get infinite--volume physics in the large--N limit. We find promising initial results, both volume reduction and the mode number method seem to work, but the effective volume of our lattices is probably still too small to reliably determine the mass anomalous dimension at the IRFP, and so results at larger N are required.
}

\FullConference{The 30th International Symposium on Lattice Field Theory\\
                 June 24 -- 29,  2012\\
                 Cairns, Australia}

\begin{document}

\section{Introduction}

Many conformal and near--conformal gauge theories have been studied non--perturbatively in the last few years, motivated largely by their possible application to dynamical models of electroweak symmetry breaking. In particular the SU(2) gauge theory with two Dirac fermions in the adjoint representation, known as Minimal Walking Technicolor (MWT)~\cite{Sannino:2004qp,Luty:2004ye}, has been the subject of many non--perturbative studies, and is one of the best understood of these conformal or near--conformal gauge theories. A particularly important quantity from a phenomenological point of view in these theories is the size of the mass anomalous dimension, which has been determined for MWT in several non--perturbative lattice studies~\cite{Bursa:2009we,DelDebbio:2010hx,DeGrand:2011qd,Catterall:2011zf}. Another measurement was recently performed using the mode number of the Dirac operator~\cite{Patella:2012da}, which gave a very precise value.

SU(N) gauge theories with adjoint Dirac fermions have also been widely studied in the context of large--N volume reduction. This is the idea that in the limit of large N, and as long as center symmetry and translational symmetry are not broken, the theory is volume independent. This means that infinite volume physics can be obtained from a small lattice, or even a single site~\cite{Eguchi:1982nm}.

Unfortunately it was quickly realised that center symmetry was spontaneously broken on small lattices in the weak coupling phase. One solution to this is to use twisted boundary conditions, which restore center symmetry at weak coupling~\cite{GonzalezArroyo:1982hz}. The original choice of twist was shown to break down at intermediate couplings for sufficiently large N, but this issue can be resolved by scaling the twist with N~\cite{GonzalezArroyo:2010ss}. Another solution is to add adjoint fermions with periodic boundary conditions~\cite{Kovtun:2003hr}, which, as long as they are sufficiently light, have been shown both perturbatively and non--perturbatively to restore center--symmetry, and hence volume independence.

In this work we attempt to determine the mass anomalous dimension of the SU(N) gauge theory with two Dirac fermions in the adjoint representation, in the limit of large N, from the mode number of the Dirac operator as done in Ref.~\cite{Patella:2012da}. Taking advantage of large--N volume reduction we do this on a very small lattice, but we should still get infinite--volume physics in the large--N limit. The large N mass anomalous dimension is also expected to be close to the N=2 case, based on the first two orders of perturbation theory which are universal and which predict that the mass anomalous dimension is independent of N.

\section{Method}
\begin{figure}
  \centering
    \includegraphics[angle=0,width=5.0cm]{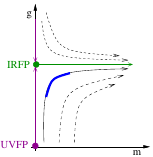}
  \caption{RG flows in a mass--deformed conformal field theory. Flow lines go from high energy scales (large eigenvalues / UV physics) to low energy (low eigenvalues / IR physics). To extract information about the IRFP we want to look at an intermediate range of eigenvalues, shown in blue.}
  \label{fig:RG}
\end{figure}

In a massless conformal field theory (CFT), the spectral density $\rho$ of the Dirac operator at small eigenvalues $\omega$ scales as~\cite{DelDebbio:2010ze} 
\begin{equation}
\rho(\omega) \propto \omega^{\frac{3-\gamma_*}{1+\gamma_*}},
\end{equation}
where $\gamma_*$ is the mass anomalous dimension at the infrared fixed point (IRFP). On the lattice the CFT is deformed by an explicit mass term, and so this relation is only approximately valid for some intermediate range of eigenvalues, which are low enough for the renormalisation flow to go close to the IRFP, but not so low that the flow has been driven away from the IRFP in the relevant mass direction, as shown schematically by the blue region in Fig.~\ref{fig:RG}.
Following Ref.~\cite{Patella:2012da}, the mode number per unit volume $\overline{\nu}(\Omega)$ of the massive hermitian Dirac operator $Q^2$ is simply the number of eigenvalues of the operator below some value $(a\Omega)^2$ divided by the volume (the volume being $\propto $N$^2L^4$ in this case),
\begin{equation}
\overline{\nu}(\Omega) = 2 \int_0^{\sqrt{\Omega^2-m^2}}\rho(\omega)\,\, d\omega,
\end{equation}
which can be integrated and written in lattice units as
\begin{equation}
a^{-4}\overline{\nu}(\Omega) \simeq a^{-4}\overline{\nu}_0 + A\left[(a\Omega)^2-(am)^2\right]^{\frac{2}{1+\gamma_*}}.
\label{eq:fit}
\end{equation}
From this one can extract $\gamma_*$ by fitting the four free parameters of this fit in some intermediate range $\overline{\Omega}_{IR}<\Omega<\overline{\Omega}_{UV}$, whilst maintaining the separation of scales on the lattice,
\begin{equation}
\frac{1}{L\sqrt{\mathrm{N}}} \ll m \ll \overline{\Omega}_{IR} < \Omega < \overline{\Omega}_{UV} \ll \frac{1}{a}.
\end{equation}

\section{Simulation Details}
We use a modified version of the HiRep~\cite{DelDebbio:2008zf} code, with the Wilson plaquette gauge action, and two Wilson fermions in the adjoint representation. We simulate with both symmetric boundary conditions ($k=0$) and the minimal symmetric twist of Ref.~\cite{GonzalezArroyo:1982hz} ($k=1$), on $2^4$ lattices with N up to 25. The use of $2^4$ lattices instead of the single site gives a combination of volume and large--N scaling, and in addition allows us to use even--odd preconditioning of the Dirac operator which significantly improves the performance of the simulations. 

In order to get continuum physics, we need to stay on the weak coupling side of the strong coupling lattice transition, which can be identified by a discontinuity in the plaquette as a function of the coupling $\lambda=\mathrm{N} g_0^2$. In addition, for volume independence to hold, we need center--symmetry to be preserved in the large--N limit, which on the lattice means the Polyakov loop modulus is zero.

\begin{figure}
  \centering
    \includegraphics[angle=270,width=8.8cm]{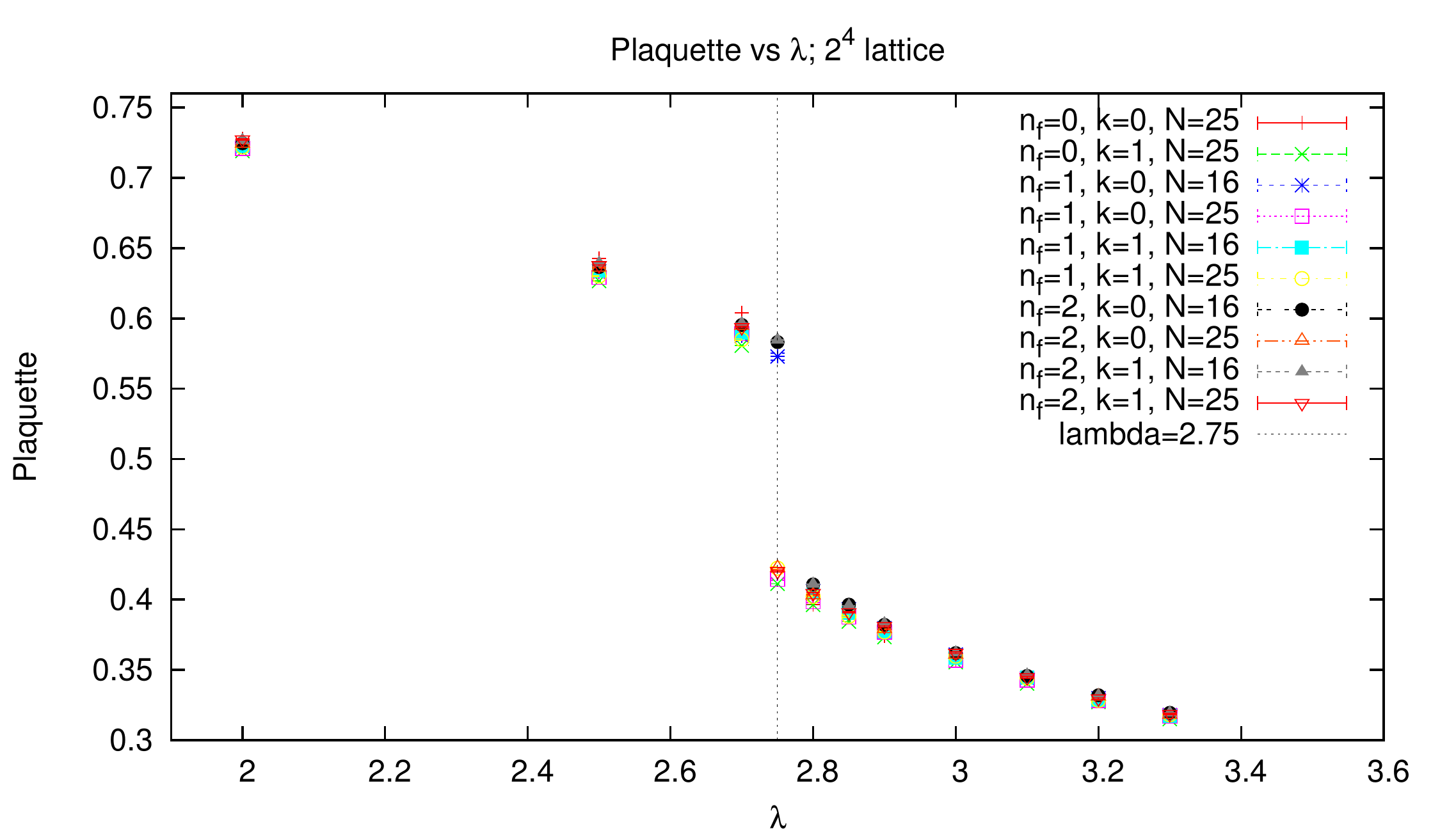}
  \caption{Plaquette as a function of $\lambda$. The strong coupling transition at $\lambda \simeq 2.75$ does not depend on twist, the number of colors $N$ or the number of flavors $n_f$.}
  \label{fig:nfX_twistX}
\end{figure}

Fig.~\ref{fig:nfX_twistX} shows the plaquette as a function of $\lambda$ for 0,1 and 2 flavours, with and without twisted boundary conditions, for N=16 and 25. The location of the strong coupling transition does not depend on the number of adjoint fermions, or on whether the boundary conditions are twisted or not. It lies at $\lambda\simeq 2.75$ in all cases, so we choose to work at $\lambda=2.70$, choosing a fairly strong coupling to minimise finite volume effects, but taking care to remain on the weak coupling side of the strong coupling transition.

\section{Results}
\begin{figure}
  \centering
    \includegraphics[angle=270,width=6.6cm]{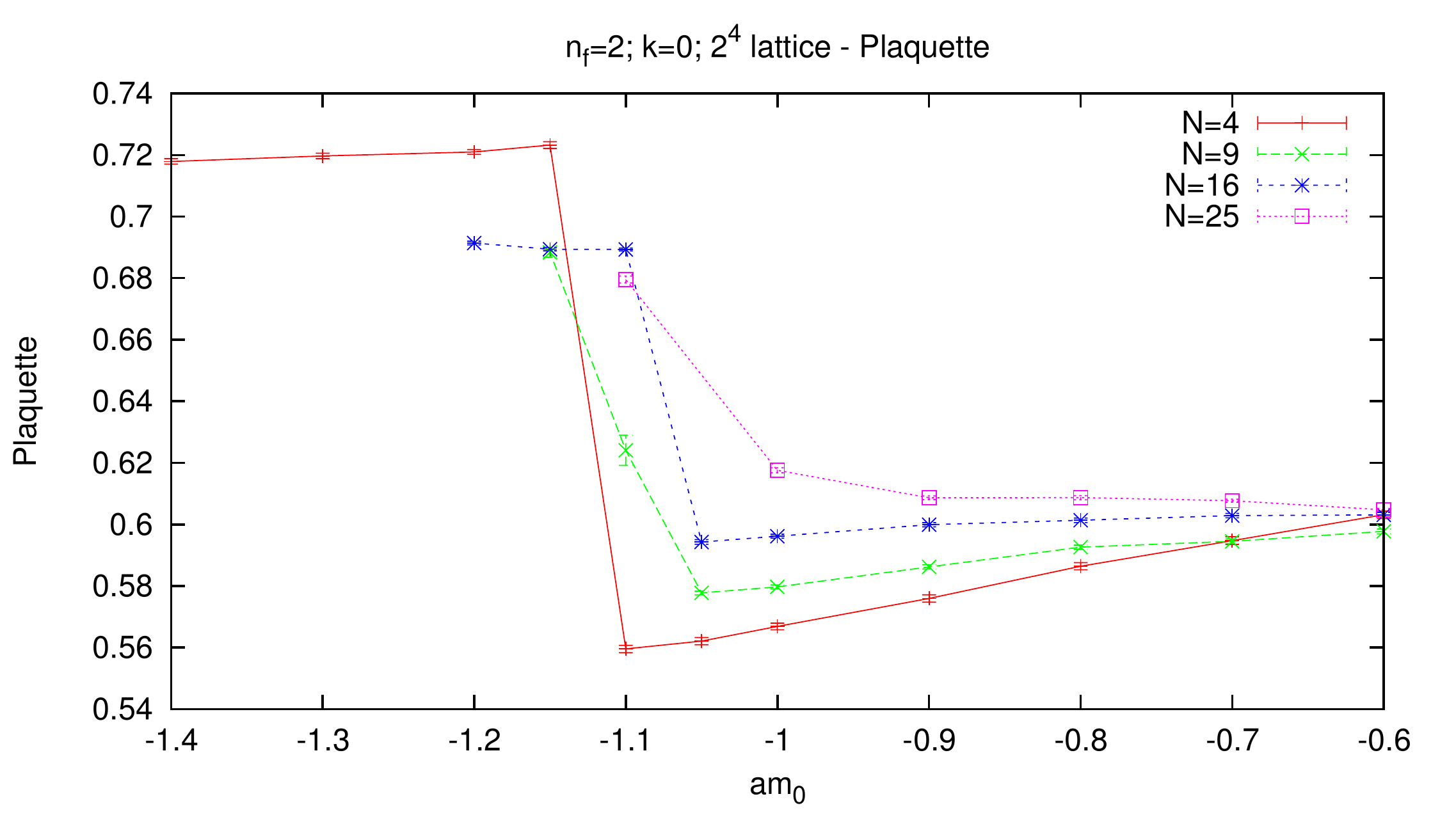}
    \includegraphics[angle=270,width=6.6cm]{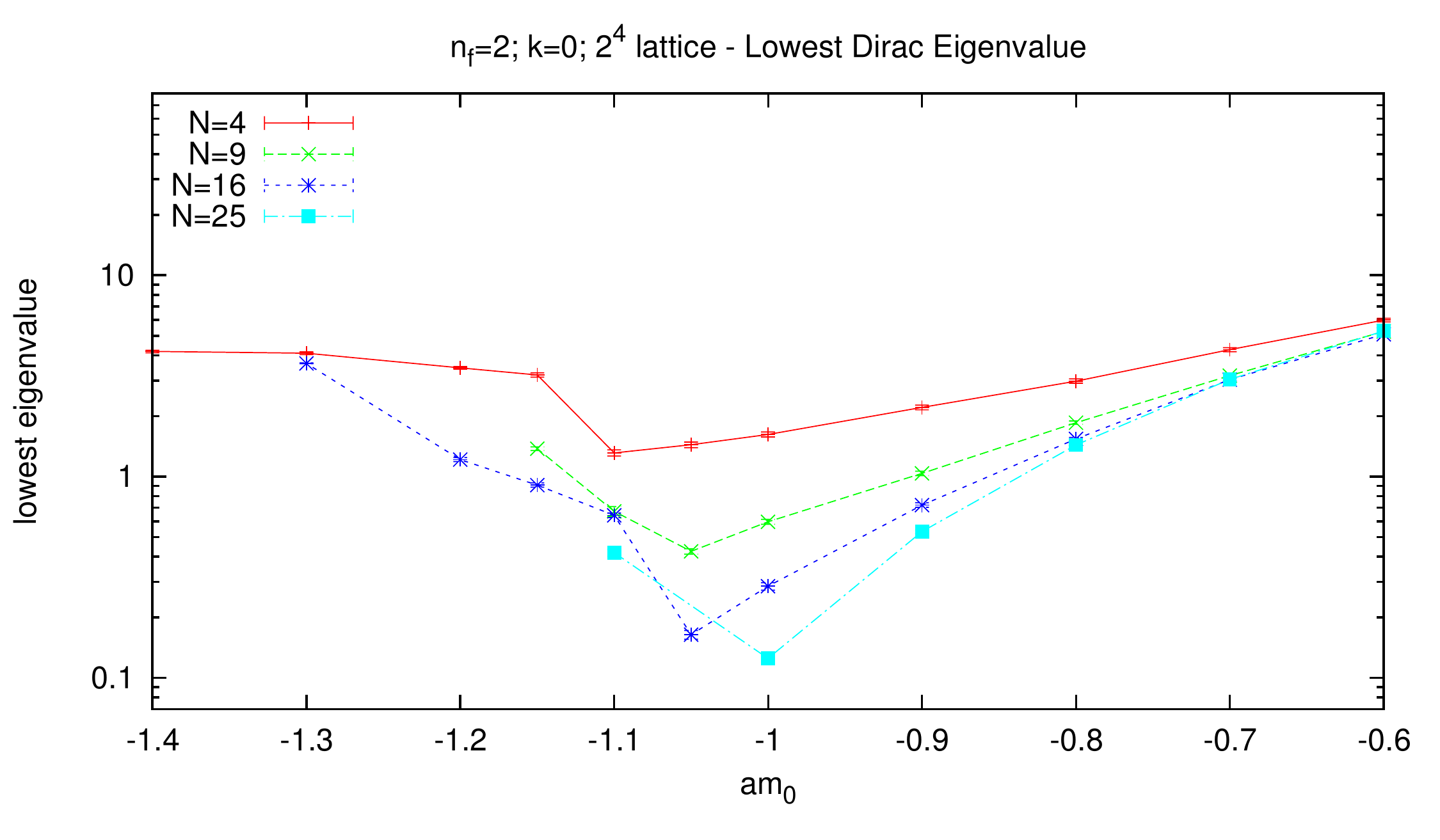}

    \includegraphics[angle=270,width=6.6cm]{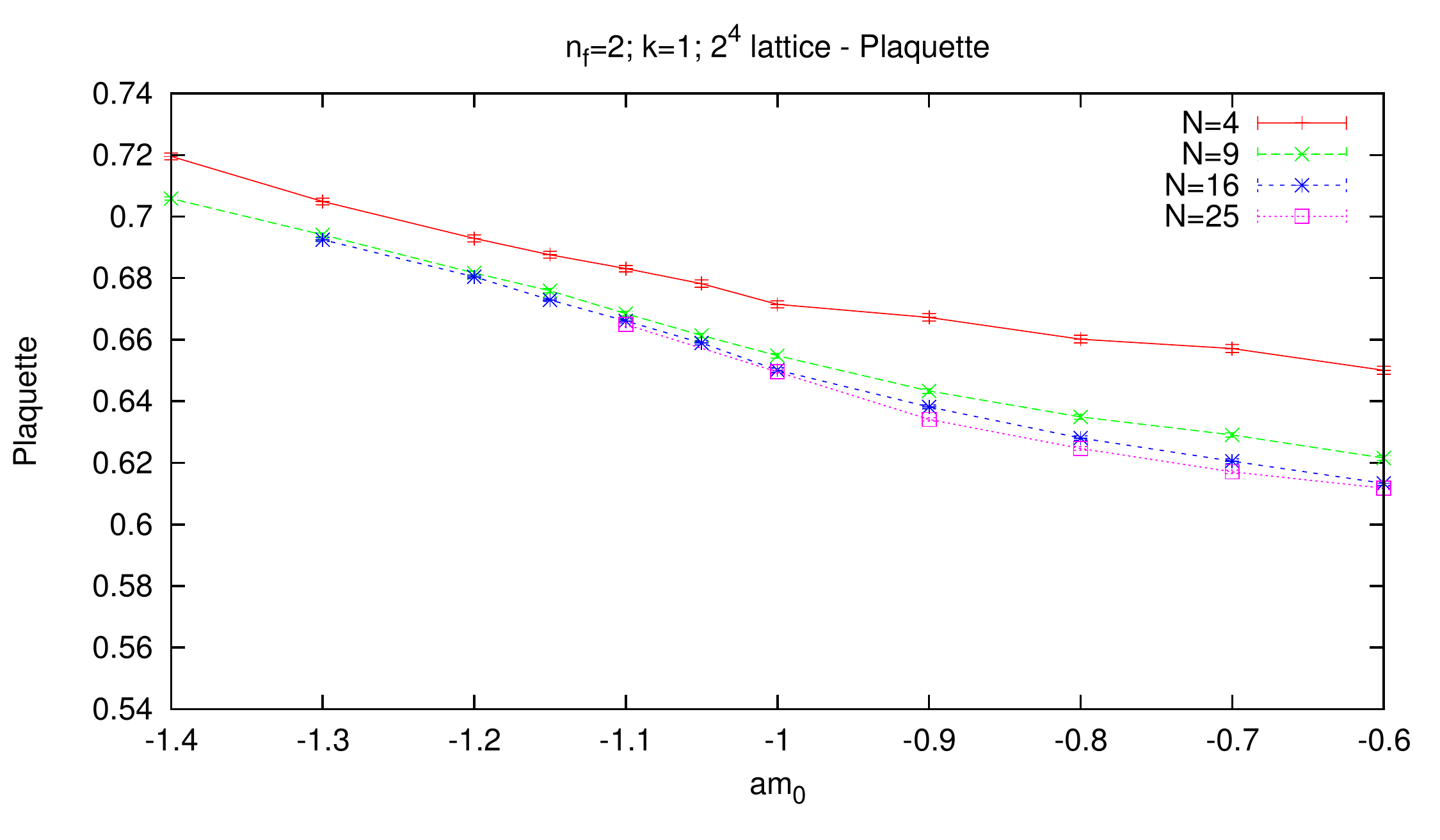}
    \includegraphics[angle=270,width=6.6cm]{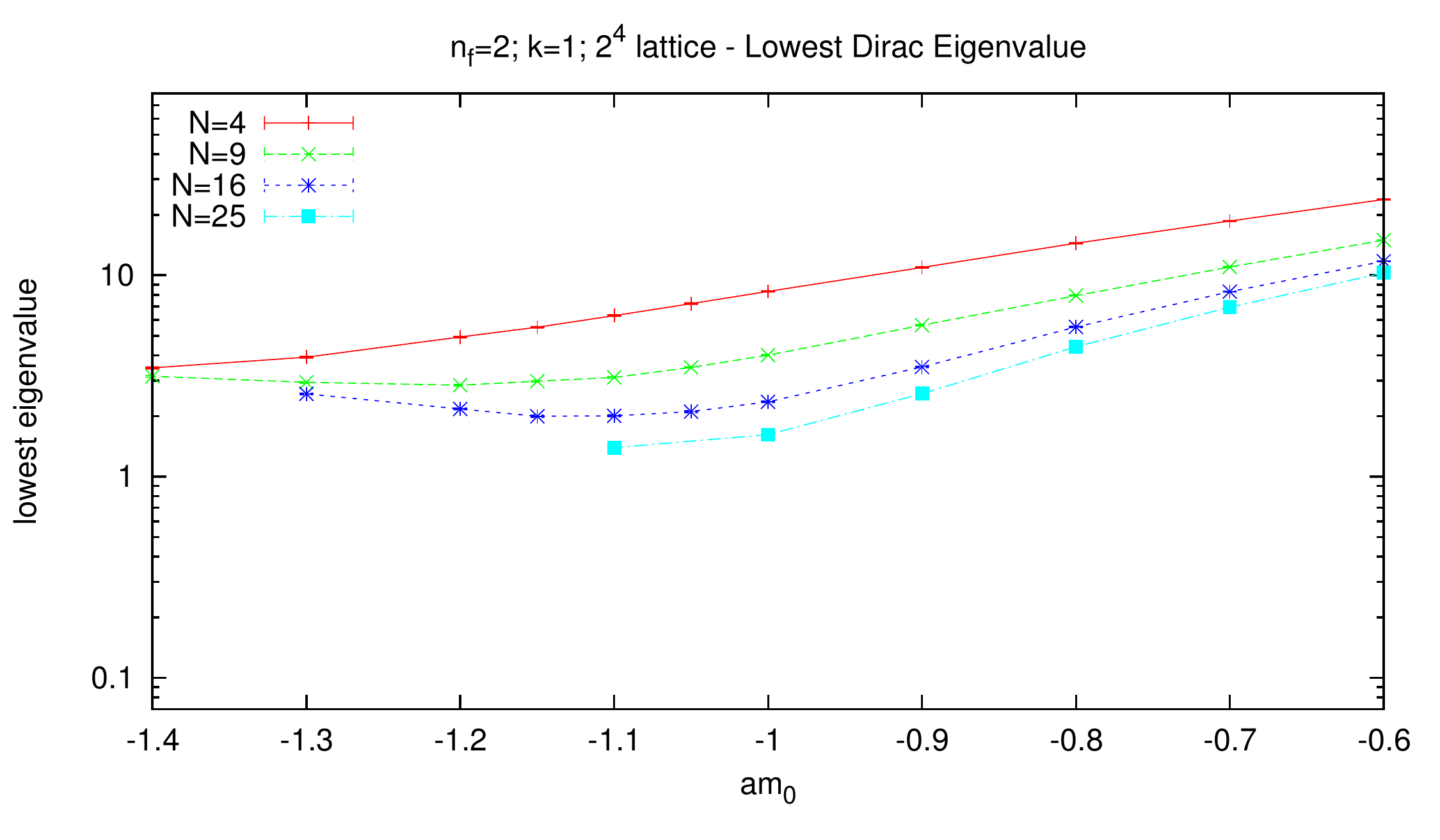}
  \caption{Plaquette and lowest eigenvalue of the Dirac operator as a function of the bare mass $am_0$. Upper plots show the untwisted $k=0$ data, lower plots show the twisted $k=1$ data.}
  \label{fig:nf2_lambda270_scan}
\end{figure}

Fig.~\ref{fig:nf2_lambda270_scan} shows the plaquette and the smallest eigenvalue of the Dirac operator, as a function of the bare mass. For the untwisted case we can identify a critical bare mass $am_0\simeq-1.05$ by a minimum in the lowest eigenvalue of the Dirac operator. At this bare mass we also see a discontinuity in the plaquette, as reported in Ref.~\cite{Bringoltz:2011by}, whose magnitude decreases as N is increased. For the twisted case the discontinuity in the plaquette disappears, which suggests the discontinuity may also disappear for large enough N in the untwisted case. The twist also increases the lowest eigenvalue of the Dirac operator, significantly improving the condition number of the operator, but the minimum still lies in the same location as for the untwisted case, at least for the larger values of N.

\begin{figure}
  \centering
    \includegraphics[angle=270,width=7.2cm]{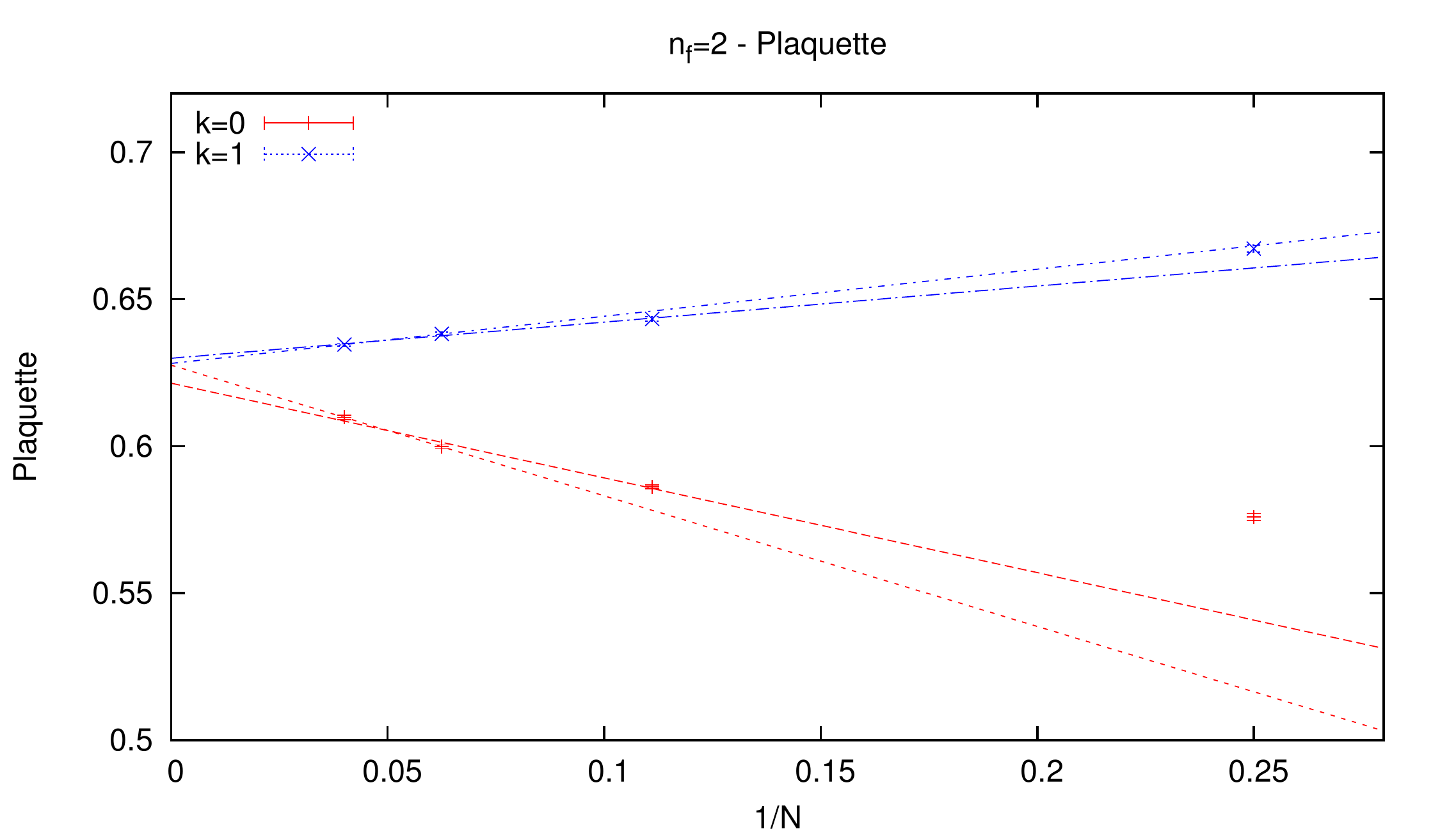}
    \includegraphics[angle=270,width=7.2cm]{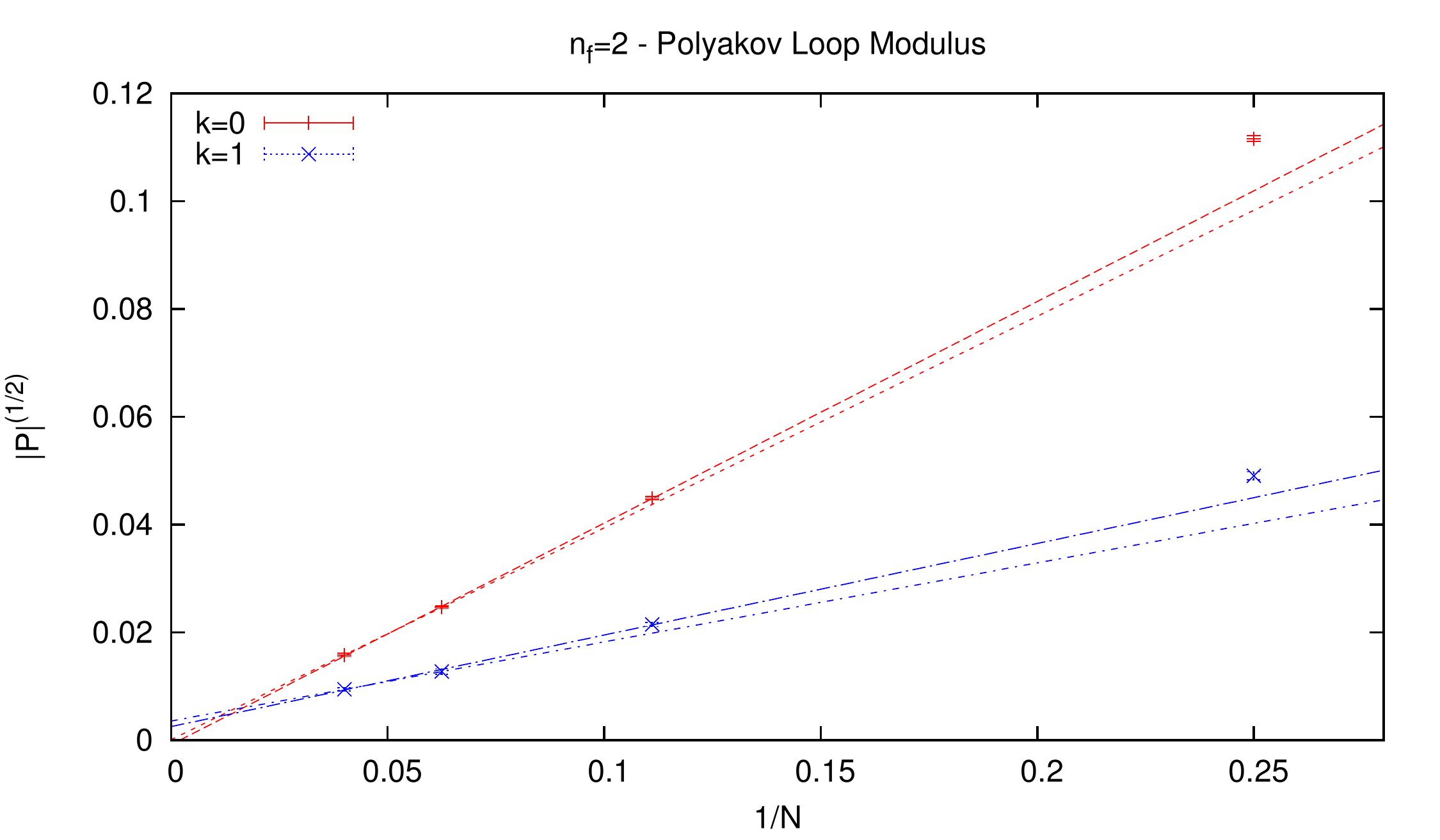}
  \caption{Plaquette and Polyakov loop modulus vs $1/N$ at $am_0=-0.90$. Two linear extrapolations shown using two or three data points. The plaquette extrapolates to a consistent value, and the Polyakov loop extrapolates to zero, indicating that center symmetry is preserved.}
  \label{fig:nf2_lambda270_plaq_poly_m090}
\end{figure}

Fig.~\ref{fig:nf2_lambda270_plaq_poly_m090} shows the plaquette and the modulus of the Polyakov loop as a function of 1/N, for both the twisted and untwisted cases at a bare mass $am_0=-0.90$. The Polyakov loop modulus goes linearly towards zero with 1/N for both the twisted and the untwisted case. For the twisted case the amount of symmetry breaking at a given N is much reduced, although the linear fit appears to go to a small but non--zero value in the large--N limit. This may be due to working at fixed $k=1$, more data at larger N and with a different value of $k$ would be needed to clarify this. For all masses the picture is qualitatively similar, and in general the twisted and untwisted plaquettes seem to linearly extrapolate to a consistent value as 1/N$\rightarrow 0$, except at the lightest mass where there is some discrepancy.

\section{Mode Number}
\begin{figure}
  \centering
    \includegraphics[angle=270,width=7.5cm]{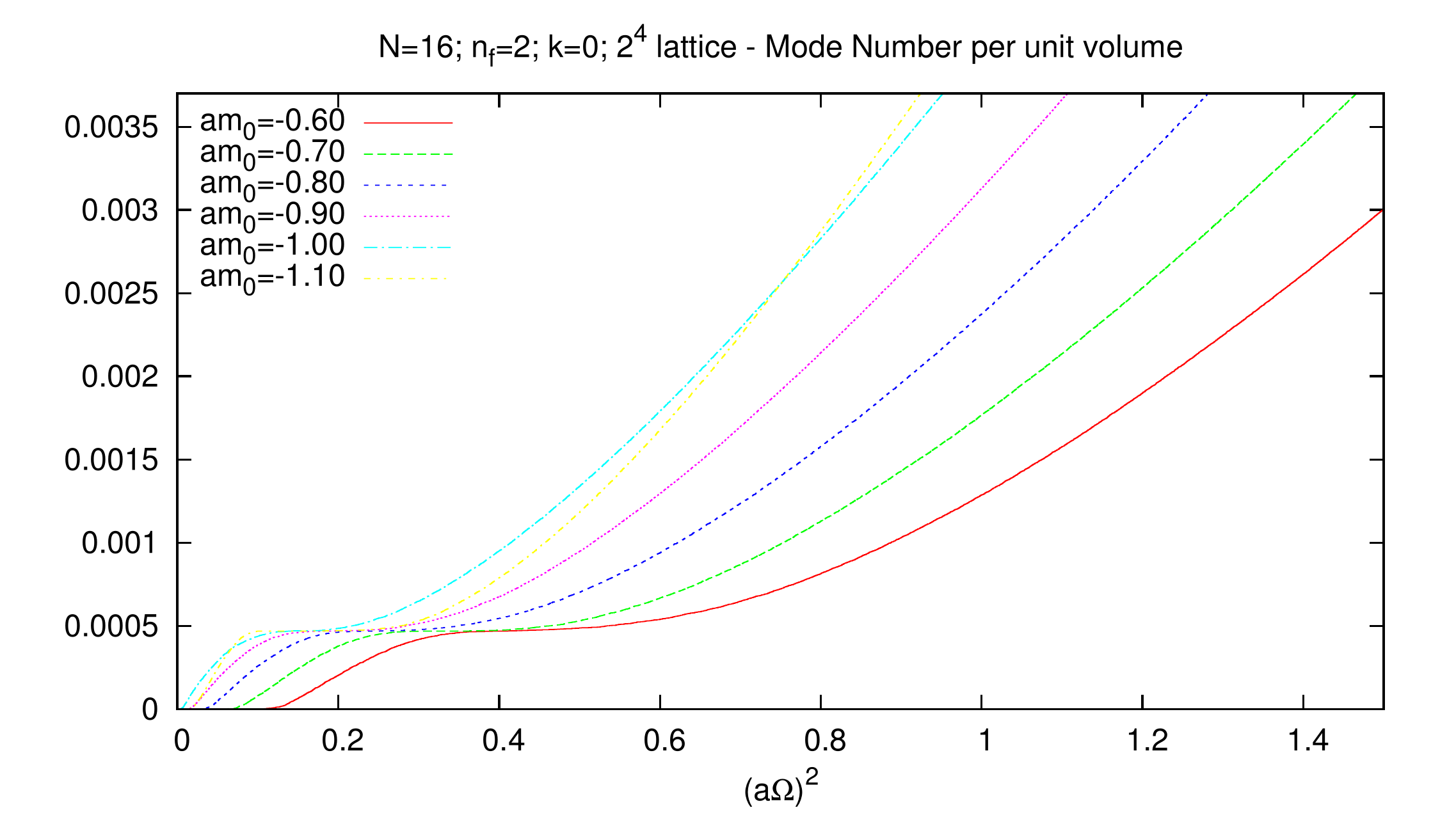}
    \includegraphics[angle=270,width=7.5cm]{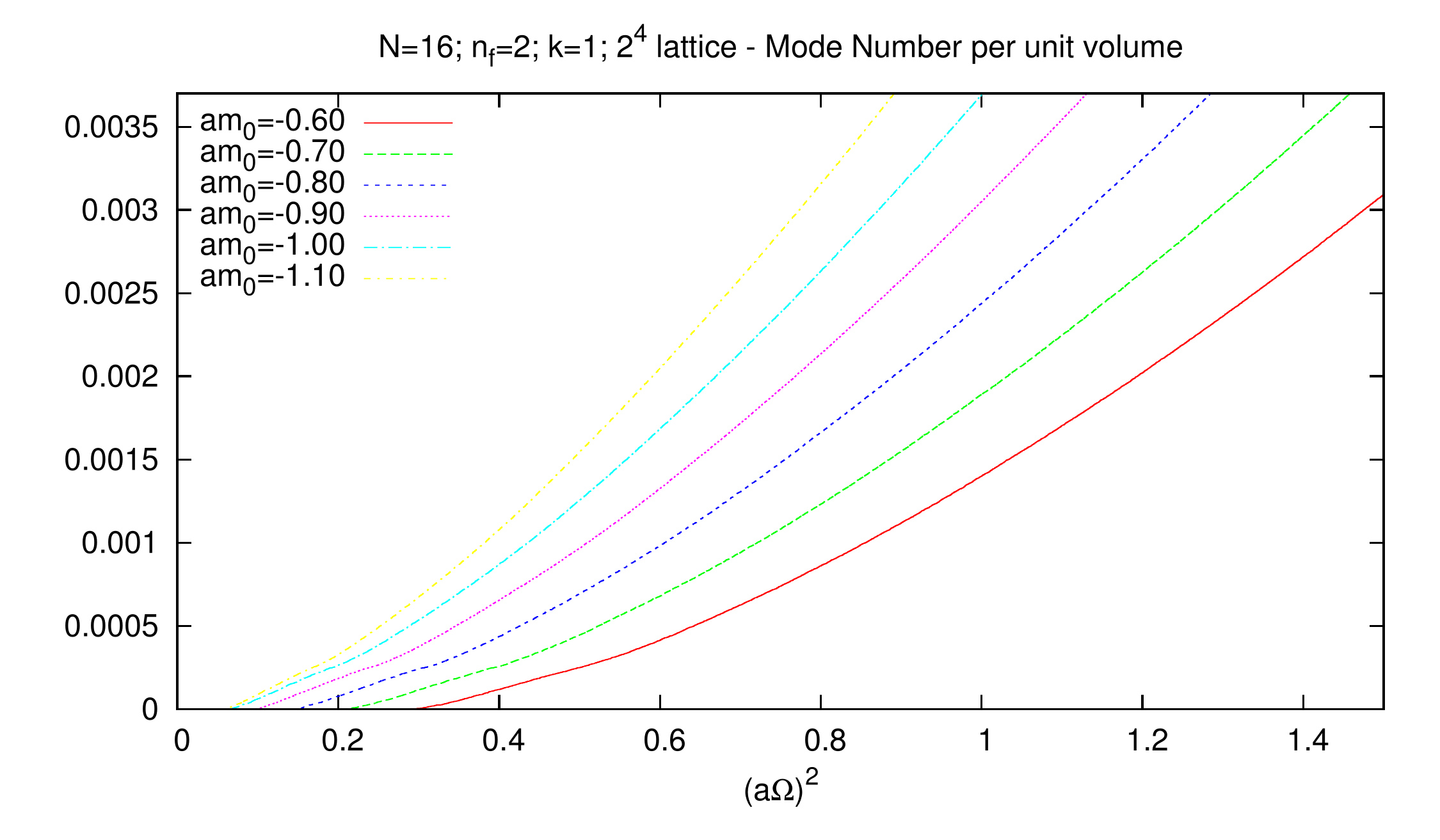}
  \caption{Mode number of Dirac operator at fixed N=16, for various bare masses $am_0$.}
  \label{fig:nf2_lambda270_eig16}
\end{figure}

Having seen that center symmetry is preserved in our simulations, and that the twisted and untwisted formulations give consistent results, we can go on to consider the mode number of the Dirac operator. Fig.~\ref{fig:nf2_lambda270_eig16} shows the Dirac mode number of Eq.~(\ref{eq:fit}) for N=16 for a range of bare masses. For the untwisted case we see the 4(N-1) would--be zero modes, which are unphysical and suppressed by $1/$N$^2$ in the large--N limit, followed by a gap, then the rest of the spectrum. As the mass is made lighter the mass gap decreases, and if we go beyond the critical bare mass then we see the mass gap increases again. For the twisted case we don't see these would--be zero modes since they are suppressed by the twist, but otherwise the picture is much the same.

\begin{figure}
  \centering
    \includegraphics[angle=270,width=7.5cm]{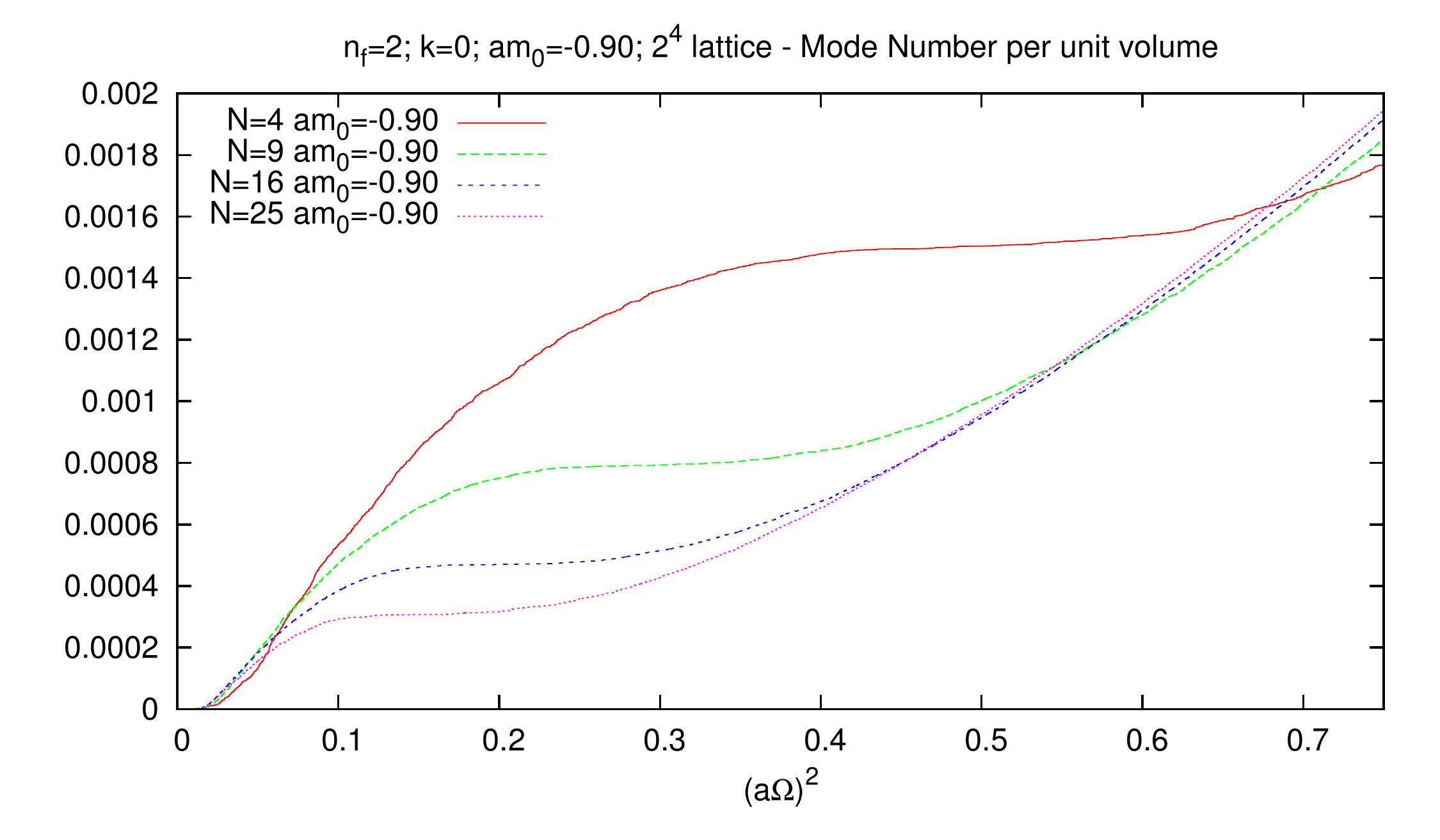}
    \includegraphics[angle=270,width=7.5cm]{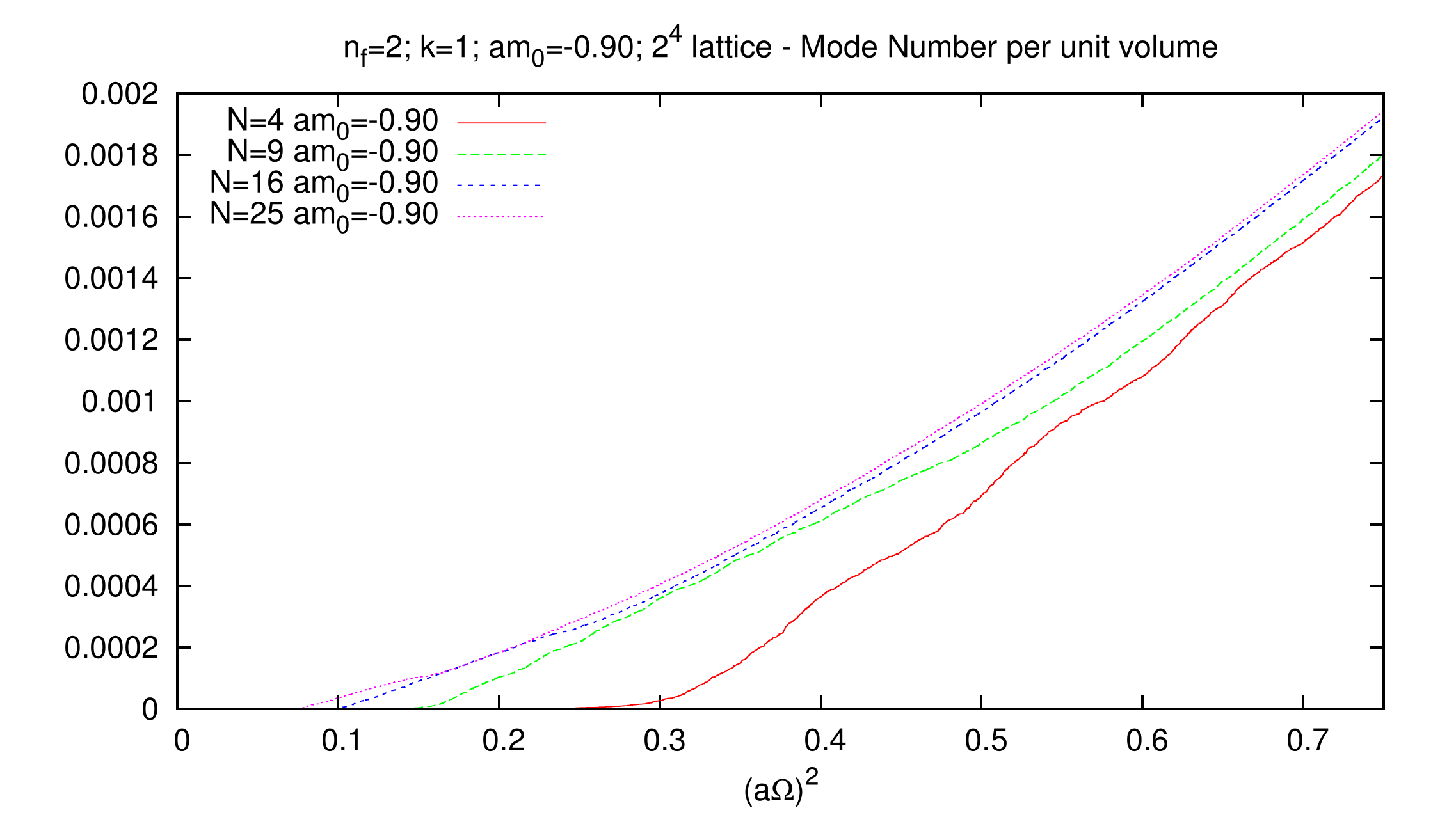}
  \caption{Mode number of Dirac operator at fixed bare mass $am_0=-0.90$, for various N.}
  \label{fig:nf2_lambda270_eig_m090}
\end{figure}

Fig.~\ref{fig:nf2_lambda270_eig_m090} shows the mode number for fixed bare mass $am_0=-0.90$, varying N. For the untwisted case, above the would--be zero modes, which are suppressed in the large N limit, there is broad agreement between different N. This makes sense, the larger eigenvalues probe a more UV region where finite volume effects are negligible, whereas the smallest eigenvalues probe the IR physics and so finite volume effects become increasingly important. The twisted mode number shows much better agreement between the different N throughout the whole range of eigenvalues.

\begin{figure}
  \centering
    \includegraphics[angle=270,width=7.5cm]{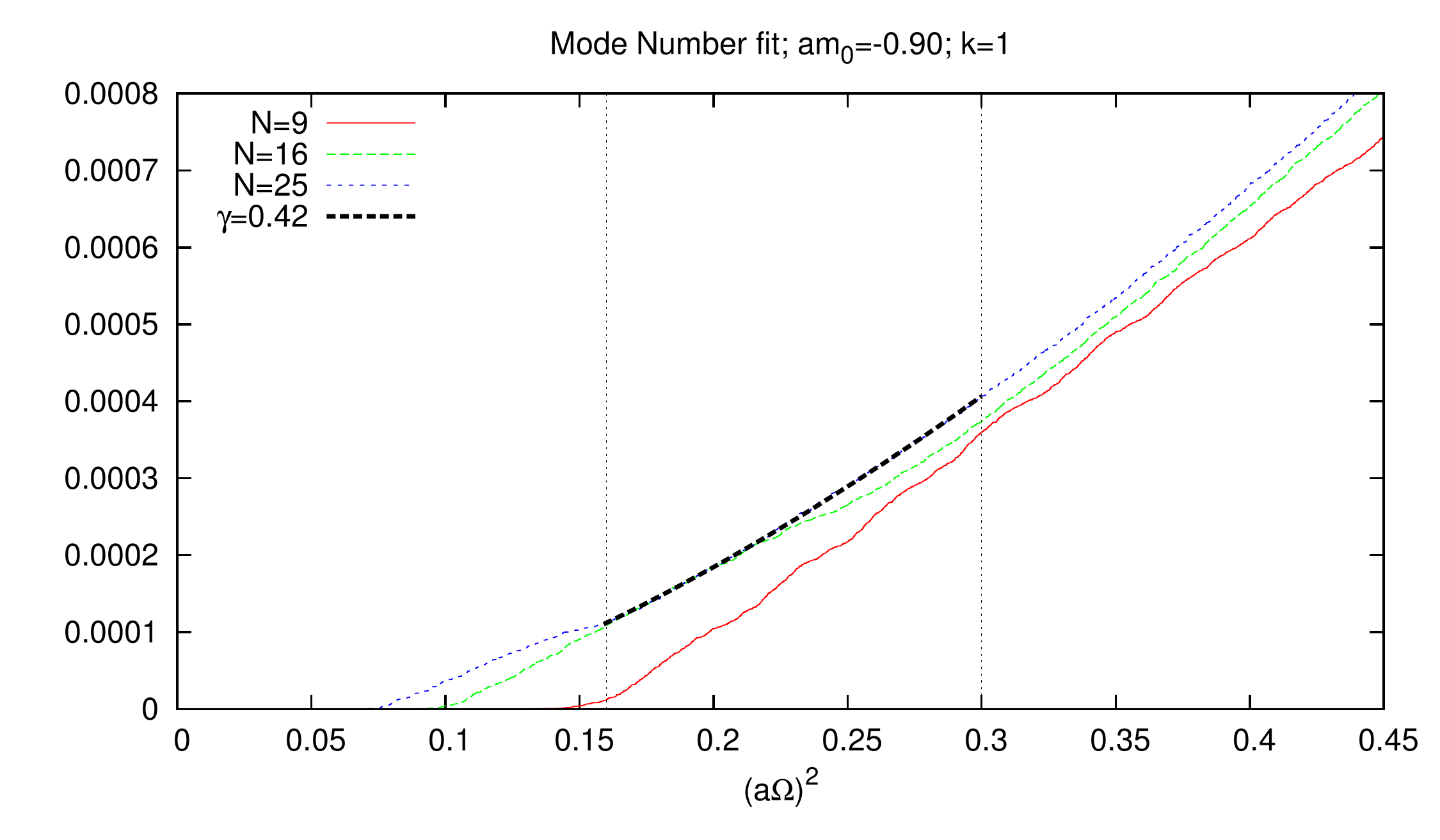}
    \includegraphics[angle=270,width=7.5cm]{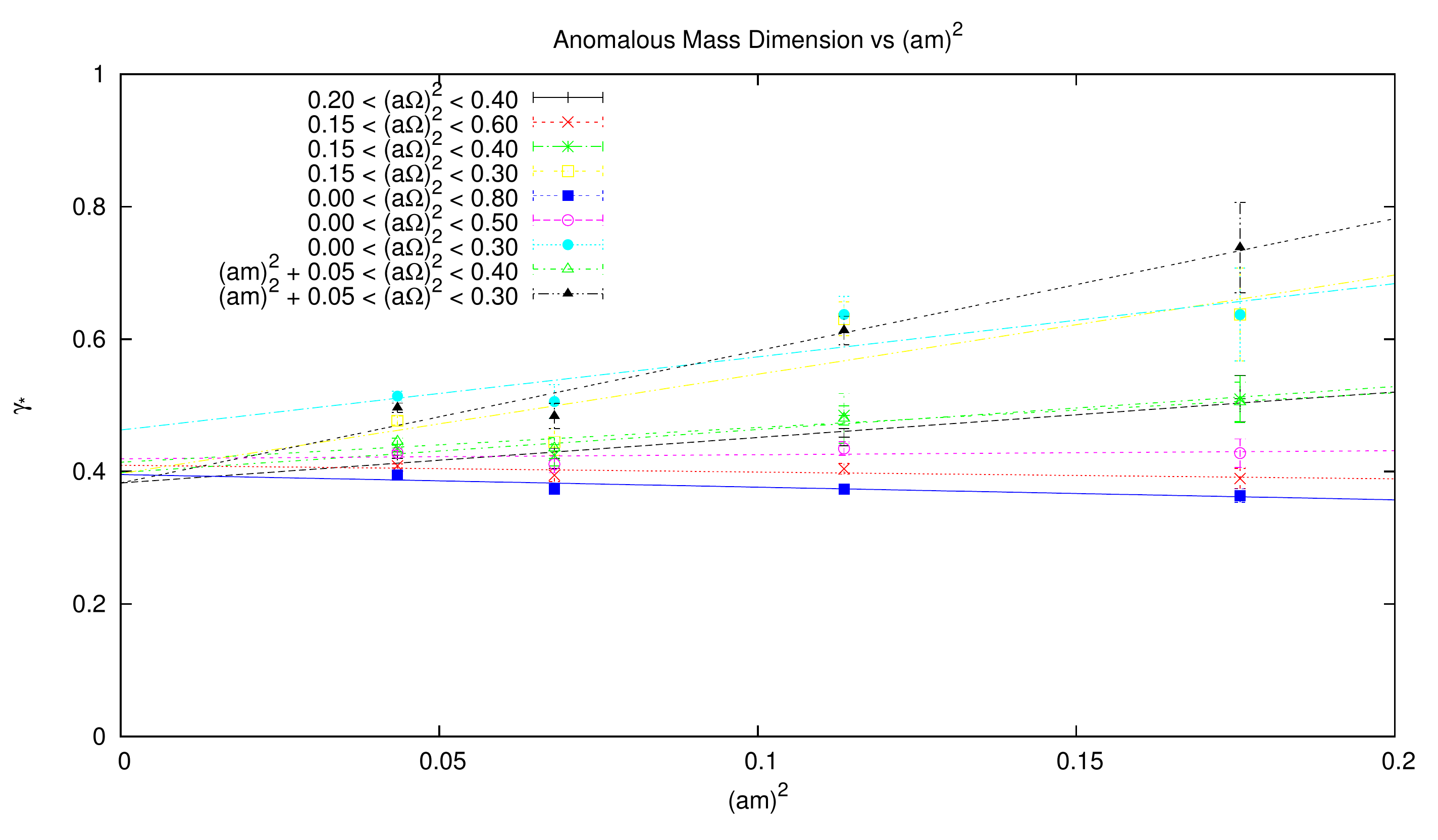}
  \caption{On the left, an example fit using $k=1$ N=25 data at $am_0=-0.90$, $0.16 < (a\Omega)^2 < 0.30$, giving $\gamma_* \sim 0.42$. On the right, different values of $\gamma_*$ are obtained by varying the fit range, and are plotted against $(am)^2$.}
  \label{fig:nf2_lambda270_eig_gamma}
\end{figure}

Fig.~\ref{fig:nf2_lambda270_eig_gamma} shows an attempt to determine $\gamma_*$ by fitting the mode number at N=25 with $k=1$ to Eq.~(\ref{eq:fit}). On the left is an example of such a fit for $am_0=-0.90$, $0.16 < (a\Omega)^2 < 0.30$, where $(am)^2$ is fixed to the smallest eigenvalue of the Dirac operator, giving $\gamma_*\sim 0.42$. On the right different values for $\gamma_*$, which were obtained in the same way but varying the fit range, are plotted against $(am)^2$, and show that there is clearly a large systematic dependence on the fit range. In addition, whatever fit range we choose, there is no guarantee that we are at large enough N and small enough mass to be close enough to the IRFP for the fit form we are using to be valid. Given this caveat, however, the extrapolation to the massless limit looks reasonably well behaved and it is encouraging to see the dependence on the fit range reduce significantly as the mass becomes lighter.

\section{Conclusion}
We find promising initial results, firstly for both the twisted and untwisted formulation we find that center symmetry is preserved, and so we should have volume independence. In addition the plaquettes from the two formulations, whilst differing significantly at finite--N, extrapolate to a consistent value in the large--N limit.
The Dirac mode number also shows broad agreement between the two formulations, at least for larger eigenvalues, and particularly in the twisted case the finite--N corrections appear to be quite small. Furthermore fits to the data give values for $\gamma_*$ that are of the same order as those found for the SU(2) theory, and as the mass is made lighter the dependence on the fit range is reduced. There remains a significant difficulty in how to determine from the data where the correct range of eigenvalues lies from which to extract the mass anomalous dimension, indeed it is possible that the values of N used are not large enough for there to exist such a range. Nevertheless the method seems to work even for our relatively small values of N, results at larger values of N are now required to further investigate this issue and allow for a reliable determination of the anomalous mass dimension.

\appendix
\section*{Acknowledgments}
We acknowledge financial support from the MCINN grants FPA2009-08785 and FPA2009-09017, the Comunidad Aut\'{o}noma de Madrid under the program HEPHACOS S2009/ESP-1473, and the European Union under Grant Agreement number PITN-GA-2009-238353 (ITN STRONGnet), and use of the IFT clusters and the SVG cluster of the Galicia Supercomputing Centre CESGA. Thanks to Margarita Garcia--Perez and Carlos Pena for interesting discussions about this work.

\end{document}